\documentclass[12pt]{article}
\usepackage{a4,amsmath,epsfig,times,euler,euscript}
\renewcommand{\baselinestretch}{1.1}
\newcommand{\Figure}[1]{Figure~\ref{#1}}

\newcommand{\Section}[1]{section~\ref{#1}}
\newcommand{\srac}[2]{{\textstyle\frac{#1}{#2}}}
\newcommand{\gof}{GOF}
\newcommand{\pdf}{PDF}
\newcommand{\td}{TD}
\newcommand{\sample}[3]{\{#1\}_{#2=1}^{#3}}
\newcommand{\sequence}[3]{(#1)_{#2=1}^{#3}}
\newcommand{\dat}{\omega}
\newcommand{\ndat}{N}
\newcommand{\ndim}{\mathit{dim}}
\newcommand{\vhi}{\varphi}
\newcommand{\Pdns}{P}
\newcommand{\Test}{T}
\newcommand{\tinf}{t_{\infty}}
\newcommand{\Ptest}{\mathcal{P}}
\newcommand{\dirac}{\delta}

\newcommand{\disc}{D}
\newcommand{\fclass}{\mathcal{H}}
\newcommand{\fmeas}{\mu}
\newcommand{\ffunc}{f}
\newcommand{\fpower}{r}
\newcommand{\dspace}{\Omega}
\newcommand{\dfamily}{\mathcal{S}}
\newcommand{\dset}{S}
\newcommand{\cmu}{c}
\newcommand{\bmu}{b}
\newcommand{\Jdet}{J}
\newcommand{\Gof}{T}
\newcommand{\Qdns}{Q}
\newcommand{\ufunc}{u}
\newcommand{\Lin}{L}
\newcommand{\nfunc}{M}
\newcommand{\weight}{\sigma}
\newcommand{\Fmeas}{\mu}
\newcommand{\cnu}{c}
\newcommand{\Gen}{G}
\newcommand{\Action}{S}
\newcommand{\imag}{\mathrm{i}}
\newcommand{\Amat}{A}

\newcommand{\nequal}{\rho}
\newcommand{\EE}{E}
\newcommand{\VV}{V}
\newcommand{\Nmeas}[1]{\langle#1\rangle_{\ndat}}

\newcommand{\Umeas}[1]{\langle#1\rangle}
\newcommand{\Exp}{E}
\newcommand{\Var}{V}
\begin{document}
%
\begin{flushright}
MZ-TH/04-05\\
\today
\end{flushright}
\vspace{\baselineskip}
\begin{center}
%
{\bf\Huge Goodness-of-fit tests in many dimensions}

\vspace{2\baselineskip}
{\Large Andr\'e van Hameren%
}

\vspace{0.25\baselineskip}
{\it\large Institut f\"ur Physik, Johannes-Gutenberg-Universit\"at,\\
         Staudinger Weg 7, D-55099 Mainz, Germany}

\vspace{0.25\baselineskip}
{\tt\large andrevh@thep.physik.uni-mainz.de}

\renewcommand{\baselinestretch}{1}
\vspace{2\baselineskip}
{\bf Abstract}\\\vspace{0.5\baselineskip}
\parbox{0.8\linewidth}{\small\hspace{15pt}%
A method is presented to construct goodness-of-fit statistics in many
dimensions for which the distribution of all possible test results in the limit
of an infinite number of data becomes Gaussian if also the number of dimensions
becomes infinite. Furthermore, an explicit example is presented, for which
this distribution as good as only depends on the expectation value and the
variance of the statistic for any dimension larger than one.
}
\end{center}
\vspace{\baselineskip}

\section{Introduction}
Goodness-of-fit (\gof) tests are designed to test the hypothesis that a sample
of data is distributed following a given probability density function (\pdf).
The sample could, for example, consist of results of a repeated experiment, and
the \pdf\ could represent the theoretical prediction for the distribution of
these results. The test consists of the evaluation of a function of the data,
the \gof\ {\em statistic\/}, and the qualification of this result using the
probability distribution of all possible results when the hypothesis is true,
the {\em test-distribution\/} (\td). Despite the consensus that \gof\ tests are
crucial for the validation of models in the scientific process, their success
is mainly restricted to one-dimensional cases, that is, to situations in which
the data-points have only one degree of freedom. The quest for \gof\ tests
useful in situations where the number $\ndim$ of dimensions is larger than one
still continues \cite{AslanZech2002,AslanZech2002-2,Raja2003}.

In the following, we will see that the difficulty with \gof\ tests in many
dimensions is to keep them {\em distribution-free\/}, that is, to construct
them such that the \td\ is independent of the \pdf.%
\footnote{That is, for {\em binning free\/} tests, which we are considering.}
We will, however, also see
how \gof\ tests can be constructed such that the asymptotic \td, in the limit
of an infinite sample size, has a Gaussian limit for $\ndim\to\infty$ for any
\pdf, so that it only depends on the expectation value and the variance of the
\gof\ statistic this limit.
Finally, we will encounter an explicit example for which the asymptotic \td\
depends, for any \pdf, as good as only on the expectation value and the
variance of the statistic for any $\ndim>1$.

\section{The structure of goodness-of-fit tests}
A \gof\ statistic is a function $\Test_{\ndat}$ of the data sample
$\sample{\dat_i}{i}{\ndat}$ constructed such that, under the hypothesis that
the data are distributed in a space $\dspace$ following the theoretical \pdf\
$\Pdns$, there is a number $\tinf$ such that
\begin{displaymath}
   \lim_{\ndat\to\infty}\Test_{\ndat}(\sample{\dat_i}{i}{\ndat})
   = \tinf
\;\;.
\end{displaymath}
The initial, na\"\i{}ve, trust in its usefulness stems from the idea that, for a
sample of finite size, the value of $\Test_{\ndat}$ should be close to $\tinf$
if the data are distributed following $\Pdns$, and that the value of
$\Test_{\ndat}$ is probably not so close to $\tinf$ if the data are not
distributed following $\Pdns$. This idea immediately leads to the question what
is ``close'', which can be answered by the {\em test-distribution\/} (\td)
\begin{equation}
   \Ptest_{\ndat}(t)
   = \int \dirac(\,t-\Test_{\ndat}(\sample{\dat_i}{i}{\ndat})\,)
     \,\prod_{i=1}^{\ndat}\Pdns(\dat_{i})d\dat_{i}
\;\;,
\label{testdistribution}\end{equation}
where each integration variable $\dat_{i}$ runs over the whole space
$\dspace$, and $\dirac$ denotes the
Dirac distribution. $\Ptest_{\ndat}$ gives the probability distribution of the
value of $\Test_{\ndat}$ under the hypothesis that the data are indeed
distributed following $\Pdns$. If it is very low at the value of
$\Test_{\ndat}$ for the empirical data, then the hypothesis that these data are
distributed following $\Pdns$ has to be rejected; it would under that
hypothesis be very improbable to get such a value. In fact, knowledge of the 
value of the number $\tinf$ is not necessary. One only needs
to know where the bulk of the \td\ is.

The evaluation of $\Test_{\ndat}(\sample{\dat_i}{i}{\ndat})$ and the
qualification of this result with $\Ptest_{\ndat}$ constitute a \gof\ test.
Notice that the \td\ is also necessary to qualify $\Test_{\ndat}$
itself: it should consist of a peak around the expectation value%
\footnote{If $\Exp(\Test_{\ndat})\neq\tinf$ then the statistic is {\em biased}.}
\begin{displaymath} 
  \Exp(\Test_{\ndat}) = \int\Test_{\ndat}(\sample{\dat_i}{i}{\ndat})
  \,\prod_{i=1}^{\ndat}\Pdns(\dat_{i})d\dat_{i}
\;\;.
\end{displaymath}
If, for example, $\Ptest_{\ndat}$ is almost flat, then the test is useless
since any data sample will lead to a value of $\Test_{\ndat}$ that is equally
probable and the test is not capable of distincting them.

\subsection{Difficulty in many dimensions}
The difficulty with the construction of \gof\ tests for $\ndim>1$ is that
it is in general very hard to calculate $\Ptest_{\ndat}$. There is a way to
avoid this, by using the the distribution from the case that $\Pdns$ is
constant. One then needs a mapping $\vhi$ of the data-points such that the
determinant of the Jacobian matrix of this mapping is equal to $\Pdns$:
\begin{displaymath}
   \left|\det\frac{\partial X_{k}(\vhi(\dat))}{\partial X_{l}(\dat)}\right|
   =\Pdns(\dat)
\;\;,
\end{displaymath}
where $X_{k}(\dat)$ is the $k$-th coordinate of data-point $\dat$.
Under the
hypothesis that the original data are distributed following $\Pdns$, the mapped
data are distributed following the uniform distribution. For $\ndim=1$, this
mapping is simply given by the integrated \pdf, or probability {\em
distribution\/} function
\begin{displaymath}
   \vhi(\dat) = \int_{-\infty}^{\dat}\Pdns(\dat')\,d\dat'
\;\;,
\end{displaymath}
since
\begin{eqnarray}
  \int\dirac(\,t-\Test_{\ndat}(\sample{\vhi(\dat_i)}{i}{\ndat})\,)
     \,\prod_{i=1}^{\ndat}\Pdns(\dat_{i})d\dat_{i}
  &=&\int\dirac(\,t-\Test_{\ndat}(\sample{\vhi_i}{i}{\ndat})\,)
     \,\prod_{i=1}^{\ndat}d\vhi_{i}
 \nonumber\\
  &=&\Ptest^{\mathrm{uniform}}_{\ndat}(t)
\;\;,
\nonumber\end{eqnarray}
where each integration variable $\vhi_i$ runs from $0$ to $1$.
$\Ptest^{\mathrm{uniform}}_{\ndat}$ is, for popular tests, known
in the limit $\ndat\to\infty$. This asymptotic distribution
$\Ptest^{\mathrm{uniform}}_{\infty}$ is assumed not to be too different from
$\Ptest^{\mathrm{uniform}}_{\ndat}$.
Tests for which this
method can be applied are called {\em distribution-free\/}.

\subsection{Crude solution}
For $\ndim>1$, finding the mapping mentioned before is in general even
more difficult than finding $\Ptest_{\ndat}$.
At least an estimate of
$\Ptest_{\ndat}$ can be found using a straightforward Monte Carlo technique:
one just has to generate `theoretical data samples' the data-points of which are
distributed following
$\Pdns$ and make a histogram of the values of $\Test_{\ndat}$ with these
samples. Depending on how accessible the analytic structure of $\Pdns$ is,
several techniques exist for generating the theoretical samples. In the
worst case that $\Pdns$ is just given as a `black box', the
Metropolis-Hastings method can be used, possibly with its efficiency improved
by techniques as suggested in \cite{Abraham1999}. Notice that one does not need
extremely many samples, since one is, for this purpose, interested in the bulk
of the distribution, not in the tails.

Even with modern computer power, however, this Monte Carlo method can become
very time consuming, especially for large $\ndat$ and large $\ndim$. In the
next section, we will see how practical \gof\ statistics for $\ndim>1$ can be
constructed for which the asymptotic \td\ can be obtained in a more efficient
way.

\section{Construction of goodness-of-fit statistics in many dimensions}
Several \gof\ statistics for the uniform distribution in many dimensions exist.
They are called {\em discrepancies\/} \cite{Tichy1997} and intensively studied
in the field of Quasi Monte Carlo integration
\cite{Niederreiter1992,QuasiMonteCarlo}, for which one uses {\em
low-discrepancy sequences\/} of multi-dimensional integration-points. These
sequences give a faster convergence than expected from the common theory of
Monte Carlo integration, because they are distributed `more uniformly'
than uniformly distributed random sequences; they give a
\gof\ that is `unacceptably good'.
When $\ndim=1$, discrepancies can be used
directly as \gof\ tests for general \pdf{}s using the
`mapping method' mentioned before, and indeed, the {\em Kolmogorov-Smirnov\/}
statistic is equivalent to the {\em $^*$-discrepancy\/}, and the {\em
Cram\'er-von Mises\/} statistic is equivalent to the
{\em $L_{2}^{*}$-discrepancy\/}.

In the following, we will have a look at the structure of discrepancies, and
we will see how they can be deformed into \gof\ statistics for general
\pdf{}s. 

\subsection{The structure of discrepancies}
Discrepancies anticipate the fact that, if a sequence
$\sample{\dat_{i}}{i}{\ndat}$ is uniformly distributed in a space $\dspace$ and
$\ndat$ becomes large, then the average of a integrable function over the
sequence should converge to the integral over $\dspace$ of the function:
\begin{displaymath}
    \Nmeas{\ffunc} \to \Umeas{\ffunc}
    \quad\textrm{for $\ndat\to\infty$}
\;\;,
\end{displaymath}
where
\begin{displaymath}
    \Nmeas{\ffunc} = \frac{1}{\ndat}\sum_{i=1}^{\ndat}\ffunc(\dat_{i})
\quad\textrm{and}\quad
    \Umeas{\ffunc} = \int_{\dspace}\ffunc(\dat)d\dat
\;\;.
\end{displaymath}
Thus a class of functions $\fclass$ and a measure $\fmeas$ on $\fclass$ are
chosen, and the discrepancy is defined as
\begin{equation}
   \disc_{\ndat} = \bigg(\;
           \int_{\fclass}|\Nmeas{\ffunc} - \Umeas{\ffunc}|^\fpower
           \,\fmeas(d\ffunc)
           \;\bigg)^{1/\fpower}
\;\;.
\label{defDisc}\end{equation}
So it is the integration error measured in a class of functions.
For example, $\fclass$ could consist of indicator functions of a family
$\dfamily$ of subsets of $\dspace$ with $\fmeas$ such that for
$\fpower\rightarrow\infty$
\begin{displaymath}
   \disc_{\ndat} = \sup_{\dset\in\dfamily}
           \bigg|\frac{1}{\ndat}\sum_{\dat_{i}\in\dset}1
                 - \int_{\dset}d\dat\bigg|
\;\;.
\end{displaymath}
In this case, the discrepancy is the maximum error made, if the volume of each
subset is estimated using $\sample{\dat_{i}}{i}{\ndat}$. Especially interesting
are the {\em quadratic\/} discrepancies \cite{vanHameren2001}, for which
$\fpower=2$, so that they are completely determined by the two-point function
of $\fmeas$:
\begin{displaymath}
   \disc_{\ndat} = \bigg(\;
           \frac{1}{\ndat^2}\sum_{i,j=1}^{\ndat}\bmu(\dat_{i},\dat_{j})
           \;\bigg)^{1/2}
\;\;,
\end{displaymath}
with
\begin{displaymath}
    \bmu(\dat_{1},\dat_{2})
    = \cmu(\dat_{1},\dat_{2})
     -\int_{\dspace}[\cmu(\dat_{1},\dat)+\cmu(\dat_{2},\dat)]\,d\dat
     +\int_{\dspace}\!\int_{\dspace}
      \cmu(\dat,\eta)\,d\dat d\eta
,
\end{displaymath}
where
\begin{displaymath}
   \cmu(\dat_1,\dat_2) 
   = \int_{\fclass}\ffunc(\dat_1)\ffunc(\dat_2)\,\fmeas(d\ffunc)
\;\;.
\end{displaymath}
So the discrepancy is the sum of the correlations of all pairs of data-points,
measured with correlation function $\bmu$. If the measure $\fmeas$ itself is
completely determined by its two-point function, it is called {\em
Gaussian\/}.

\subsection{From discrepancies to \gof\ statistics}
Discrepancies are usually constructed in order to test the uniformity of
sequences in a $\ndim$-dimensional hyper-cube $[0,1)^{\ndim}$. We are
interested in more general cases, in which we want to test whether a sample
$\sample{\dat_{i}}{i}{\ndat}$ of data-points is distributed in a space
$\dspace$ following a given \pdf\ $\Pdns$. We will assume that there exists an
invertible mapping $\vhi$ which maps these data-points onto points
$\vhi(\dat_{i})\in[0,1)^{\ndim}$ and for which the determinant $\Jdet$ of the
Jacobian matrix is known. The hypothesis dictates that the mapped points are
distributed in the hyper-cube following
$(\Pdns\circ\vhi^{-1})/(\Jdet\circ\vhi^{-1})$. We will denote this \pdf\ by
$\Pdns$ itself from now on, the sample of mapped data-points by
$\sample{\dat_{i}}{i}{\ndat}$ and the hyper-cube by $\dspace$.

We want to use the idea, introduced before, to analise a data sample
$\sample{\dat_{i}}{i}{\ndat}$ by looking at $\Nmeas{\ffunc}$ for different
functions $\ffunc$. We will just have to keep in mind that,
if $\sample{\dat_{i}}{i}{\ndat}$ is distributed following $\Pdns$, then
\begin{displaymath}
    \Nmeas{\ffunc} \to \Umeas{\ffunc\Pdns}
    \quad\textrm{for $\ndat\to\infty$}
\;\;,
\end{displaymath}
where `$\ffunc\Pdns$' denotes point-wise multiplication. This and the
definition of the discrepancies lead us to define the statistic
\begin{displaymath}
   \Gof_{\ndat} = \bigg(\;
          \int_{\fclass}|  \Nmeas{\ffunc\Qdns} 
                         - \Umeas{\ffunc\Qdns\Pdns}|^\fpower
          \,\fmeas(d\ffunc)
          \;\bigg)^{1/\fpower}
\;\;,
\end{displaymath}
where we inserted the function $\Qdns$ for flexibility.
It could be absorbed in the definition of $\fmeas$, but we prefer this
formulation, in which we can stick to known examples for $\fmeas$.
We will see later on that the ideal choice for $\Qdns$ is
\begin{equation}
   \Qdns = 1/\sqrt{\Pdns}
\;\;.
\end{equation}
We want to focus on the quadratic discrepancies for which $\fmeas$
is Gaussian from now on.
Like in \cite{vanHameren2001}, we shall define the statistic itself as an
average case complexity, and not as a square-root of an average:
\begin{eqnarray}
   \Gof_{\ndat}
    &=& \ndat\int_{\fclass}|  \Nmeas{\ffunc\Qdns} 
                          - \Umeas{\ffunc\Qdns\Pdns}|^2
      \,\fmeas(d\ffunc)
\label{defgof}\\      
    &=& \frac{1}{\ndat}\sum_{i,j=1}^{\ndat}\Qdns\cmu(\dat_{i},\dat_{j})
      - 2\int_{\dspace}\bigg(\;\sum_{i=1}^{\ndat}
                      \Qdns\cmu(\dat_{i},\dat)\;\bigg)\Pdns(\dat)d\dat
\nonumber\\ &&\hspace{130pt}
      - \ndat\int_{\dspace}\!\int_{\dspace}
               \Qdns\cmu(\dat,\eta)\,\Pdns(\dat)\Pdns(\eta)d\dat d\eta
\;\;,
\label{defgof1}\end{eqnarray}
where
\begin{equation}
  \Qdns\cmu(\dat_{1},\dat_{2})
  = \Qdns(\dat_{1})\Qdns(\dat_{2})\,\cmu(\dat_{1},\dat_{2})
\;\;.
\end{equation}
The reason for the extra factor $\ndat$ becomes clear when we calculate the 
expectation value of $\Gof_{\ndat}$.
Assuming that the data-points are distributed independently following $\Pdns$,
it is given by
\begin{displaymath}
   \Exp(\Gof_{\ndat})
   = \int_{\dspace}\Qdns\cmu(\dat,\dat)\,\Pdns(\dat)d\dat
   - \int_{\dspace}\!\int_{\dspace}
           \Qdns\cmu(\dat_1,\dat_2)\,\Pdns(\dat_1)\Pdns(\dat_2)d\dat_1d\dat_2
\;\;.
\end{displaymath}
So it is independent of $\ndat$ and the statistic is not biased. In order to
write down the variance, we shorten the notation such that the
expectation value can be written as
\begin{equation}
   \Exp(\Gof_{\ndat})
   =   \Umeas{\Qdns\cmu_{1,1}\Pdns_{1}}
     - \Umeas{\Qdns\cmu_{1,2}\Pdns_{1}\Pdns_{2}}
\;\;,
\label{expectationvalue}\end{equation}
and the variance is given by
\begin{eqnarray}
  \Var(\Gof_{\ndat})
   = \bigg(1-\frac{2}{\ndat}\bigg)\bigg(
    &&\Umeas{\,( \Qdns\cmu_{1,2}\Qdns\cmu_{1,2}
                     +\Qdns\cmu_{1,1}\Qdns\cmu_{1,2})  \Pdns_{1}\Pdns_{2}\,}
\nonumber\\
 \quad&-&\Umeas{\,(  \Qdns\cmu_{1,1}\Qdns\cmu_{2,3}
                   +4\Qdns\cmu_{1,2}\Qdns\cmu_{2,3})
                                              \Pdns_{1}\Pdns_{2}\Pdns_{3}\,}
\nonumber\\
 \quad&+& 3\Umeas{\,\Qdns\cmu_{1,2}\Qdns\cmu_{3,4}
                                       \Pdns_{1}\Pdns_{2}\Pdns_{3}\Pdns_{4}\,}
 \hspace{70pt}\bigg)
\;\;.
\label{variance}\end{eqnarray}

Notice that the formulation with Gaussian measures on the function class
corresponds to a natural interpretation of the average of a square: given a
sequence $\sequence{\ufunc_{n}}{n}{\nfunc}$ of functions, a sequence
$\sequence{\weight_{n}^2}{n}{\nfunc}$ of positive weights and a linear
operation $\Lin$, we have
\begin{displaymath}
   \sum_{n=1}^{\nfunc}\weight_{n}^2\Lin(\ufunc_{n})^2
   = \int\Lin\bigg(\,\sum_{n=1}^{\nfunc}x_{n}\ufunc_{n}\,\bigg)^2
     \,\exp\bigg(-\sum_{n=1}^{\nfunc}\frac{x_{n}^2}{2\weight_{n}^2}
           \bigg)\,\prod_{n=1}^{\nfunc}\frac{dx_i}{\sqrt{2\pi\weight_{n}^2}}
\;\;,
\end{displaymath}
where the $x_{n}$-integrals run from $-\infty$ to $\infty$.
So the square averaged over the sequence 
$\sequence{\ufunc_{n}}{n}{\nfunc}$ and weighted with 
$\sequence{\weight_{n}^2}{n}{\nfunc}$
is equal to the square averaged
over the class of functions that can be written as linear combination of
$\sequence{\ufunc_{n}}{n}{\nfunc}$ measured with Gaussian weights with 
widths $\sequence{\weight_{n}}{n}{\nfunc}$. In the formulation of the
statistic in terms of the two-point function this means that
$\sequence{\ufunc_{n},\weight_{n}^2}{n}{\nfunc}$ gives its spectral 
decomposition:
\begin{equation}
   \cmu(\dat_{1},\dat_{2})
   = \sum_{n=1}^{\nfunc}\weight_{n}^2\ufunc_{n}(\dat_{1})\ufunc_{n}(\dat_{2})
\;\;.
\label{spectral}\end{equation}
The sequence $\sequence{\ufunc_{n}}{n}{\nfunc}$ usually consists of an
orthonormal basis, and several examples of decompositions
$\sequence{\ufunc_{n},\weight_{n}^2}{n}{\nfunc}$ can be found in
\cite{vanHameren2001}, including cases with $\nfunc=\infty$. 
One can also find the famous $\chi^2$-statistic interpreted in this way there,
with $\sequence{\ufunc_{n}}{n}{\nfunc}$ a set of indicator functions of
non-overlapping subsets of $\dspace$, and $\weight_{n}^2=1/\Umeas{\ufunc_{n}}$.

A closer look at formula (\ref{defgof1}) for the \gof\ statistic reveals that
it is highly impractical for the estimation of the \td\ with the
Monte Carlo method, firstly because it is quadratic in the number of
data-points and secondly because a $\ndim$-dimensional integral has to be
calculated for each data sample.%
\footnote{The $2\ndim$-dimensional integral
does not depend on the data sample, and has to be calculated only once.} One
such integral evaluation can be performed within acceptable time-scale using
Monte Carlo integration techniques, by generating integration-points $\dat$
distributed following $\Pdns$ and calculating the average of
$\sum_{i=1}^{\ndat}\Qdns\cmu(\dat_{i},\dat)$. In order to make an estimate of
the \td\ with a histogram, however, one would have to calculate in
the order of a thousand of such integrals.

Fortunately, the precise definition of the statistic, or more explicitly the
spectral decomposition of the two-point function, can be chosen such that the
asymptotic \td\ $\Ptest_{\infty}$ becomes Gaussian for $\ndim>\infty$, as we
will see in the next section. This indicates that, for large $\ndim$,
$\Ptest_{\infty}$ only depends on the expectation value and the variance of the
statistic. In \Section{examples}, we will see an explicit example for which
$\Pdns$ influences $\Ptest_{\infty}$ as good as only through the expectation
value and the variance for any $\ndim>1$, even before  $\Ptest_{\infty}$ looks
like a Gaussian.  So instead of thousands of $\ndim$-dimensional integrals for
a histogram, one only has to calculate a $\ndim$, two $2\ndim$, a $3\ndim$ and
a $4\ndim$-dimensional integral for the expectation value
(\ref{expectationvalue}) and the variance (\ref{variance}).

\section{Calculation of the asymptotic test-distribution}
We approach the calculation of $\Ptest_{\ndat}$ through its moment generating
function
\begin{displaymath}
   \Gen_{\ndat}(z) = \Exp(\,e^{z\Gof_{\ndat}}\,)
\;\;.
\end{displaymath}
$\Ptest_{\ndat}$ can be recovered form $\Gen_{\ndat}$ by the
inverse Laplace transformation
\begin{equation}
   \Ptest_{\ndat}(t) 
   = \int_{\Gamma}\frac{dz}{2\pi\imag}
     \,\exp(\,\Action(t;z)\,)
\quad,\quad
   \Action(t;z) = \log\Gen_{\ndat}(z) - tz
\;\;,
\label{Laplace}\end{equation}
where $\Gamma$ runs from $-\imag\infty$ to $\imag\infty$ on the left side of
any singularity of $\Gen_{\ndat}$.
%
The analysis of $\Gen_{\ndat}$ can be simplified by 
the observation that the statistic (\ref{defgof}) 
does not change if we replace
\begin{displaymath}
   \ufunc_{n} \leftarrow \ufunc_{n} 
                       - \frac{1}{\Qdns}\Umeas{\ufunc_{n}\Qdns\Pdns}
\end{displaymath}
in the spectral decomposition,
since $\Lin(\ufunc_{n})=\Nmeas{\ufunc_{n}\Qdns}-\Umeas{\ufunc_{n}\Qdns\Pdns}$
is invariant (remember that $\Umeas{\Pdns}=1$). In other words, (\ref{defgof})
with $\fmeas$ Gaussian and two-point function (\ref{spectral}) is
equivalent to
\begin{equation}
   \Gof_{\ndat}
    = \ndat\int_{\fclass}\Nmeas{\ffunc\Qdns}^2\,\Fmeas(d\ffunc)
    = \frac{1}{\ndat}\sum_{i,j=1}^{\ndat}\Qdns(\dat_{i})\Qdns(\dat_{j})
                          \cnu(\dat_{i},\dat_{j})
\;\;,
\label{defgof2}\end{equation}
with $\Fmeas$ Gaussian and two-point function 
\begin{equation}
   \cnu(\dat_{1},\dat_{2})
   = \sum_{n=1}^{\nfunc}\weight_{n}^2
     \bigg(\ufunc_{n}(\dat_{1})
           - \frac{\Umeas{\ufunc_{n}\Qdns\Pdns}}{\Qdns(\dat_{1})}\bigg)
     \bigg(\ufunc_{n}(\dat_{2})
           - \frac{\Umeas{\ufunc_{n}\Qdns\Pdns}}{\Qdns(\dat_{2})}\bigg)
\;\;.
\label{twopoint2}\end{equation}
With this decomposition, we can put $\Umeas{\ffunc\Qdns\Pdns}$
equal to zero under the measure.
We continue in the spirit of \cite{vanHamerenKleiss,vanHameren2001}, and write
\begin{displaymath}
   \Gof_{\ndat} = \int_{\dspace}\!\int_{\dspace}
                  \cnu(\dat,\eta)
                  \dirac_{\ndat}(\dat)\dirac_{\ndat}(\eta)\,d\dat d\eta
\;\;,
\end{displaymath}
where
\begin{displaymath}
   \dirac_{\ndat}(\dat) 
   = \frac{\Qdns(\dat)}{\sqrt{\ndat}}\sum_{i=1}^{\ndat}\dirac(\dat_{i}-\dat)
\;\;,
\end{displaymath}
so that, using Gaussian integration rules, we find that
\begin{displaymath}
  e^{z\Gof_{\ndat}}
  =\int_{\fclass}e^{\sqrt{2z}\,\Umeas{\ffunc\dirac_{\ndat}}}
   \,\Fmeas(d\ffunc)
  =\int_{\fclass}\bigg(\;\prod_{i=1}^{\ndat}
                 e^{\sqrt{2z/\ndat}\,\ffunc(\dat_{i})\Qdns(\dat_{i})}
                 \;\bigg)\Fmeas(d\ffunc)
\;\;,
\end{displaymath}
and
\begin{displaymath}
  \Gen_{\ndat}(z) = \Exp(\,e^{z\Gof_{\ndat}}\,)
  =\int_{\fclass}
     \Umeas{\,\Pdns\,e^{\sqrt{2z/\ndat}\,\ffunc\Qdns}\,}^{\ndat}
   \,\Fmeas(d\ffunc)
\;\;.
\end{displaymath}
We shall restrict ourselves to the asymptotic distribution for 
$\ndat\to\infty$ from now on. We find
\begin{displaymath}
  \Gen_{\infty}(z)
  =\lim_{\ndat\to\infty}\Gen_{\ndat}(z)
  =\int_{\fclass}
     e^{z\Umeas{\ffunc^2\Qdns^2\Pdns}}\,\Fmeas(d\ffunc)
\;\;,
\end{displaymath}
where we used the fact that
that $\Umeas{\ffunc\Qdns\Pdns}$ can
be taken equal to zero under the measure. Substituting
\begin{displaymath}
   \ffunc(\dat)
   = \sum_{n=1}^{\nfunc}x_{n}\bigg(\ufunc_{n}(\dat)
           - \frac{\Umeas{\ufunc_{n}\Qdns\Pdns}}{\Qdns(\dat)}\bigg)
\quad\textrm{and}\quad
   \Fmeas(d\ffunc) 
   = \prod_{n=1}^{\nfunc}e^{-\frac{x^2}{2\weight_{n}^2}}
                         \frac{dx_{n}}{\sqrt{2\pi\weight_{n}^2}}
\;\;
\end{displaymath}
and applying well known Gaussian integration rules, we find
\begin{equation}
   \Gen_{\infty}(z)
   = \det(1-2z\Amat)^{-1/2}
\;\;,
\label{Geninfty}\end{equation}
with
\begin{displaymath}
   \Amat_{n,m}
    =  \weight_{n}\weight_{m}\Umeas{\ufunc_{n}\ufunc_{m}\Qdns^2\Pdns}
      -\weight_{n}\Umeas{\ufunc_{n}\Qdns\Pdns}
       \weight_{m}\Umeas{\ufunc_{m}\Qdns\Pdns}
\;\;.
\end{displaymath}
The asymptotic generating function is now determined up to the positions of
its singularities, which can directly be written in terms of the eigenvalues
$\sequence{\lambda_{n}}{n}{\nfunc}$ of $\Amat$, since
\begin{equation}
   \Gen_{\infty}(z)
   = \bigg(\;\prod_{n=1}^{\nfunc}(1-2z\lambda_{n})\;\bigg)^{-1/2}
\;\;.
\label{Geninfty1}\end{equation}
Another way to see how the eigenvalues affect the shape of the 
\td\ is by considering the cumulants, which are generated by the
logarithm of the generating function:
\begin{displaymath}
   \frac{d^k\log\Gen_{\infty}}{dz^k}(z=0)
   = 2^{k-1}(k-1)!\sum_{n=1}^{\nfunc}\lambda_{n}^k
\;\;.
\end{displaymath}
If $\Amat$ would consist only of a diagonal term plus a diadic term, then the
access to its eigenvalues would be relatively easy. Having in mind that the
functions $\ufunc_{n}$ are orthonormal, this can be achieved by the choice
\begin{displaymath}
   \Qdns = 1/\sqrt{\Pdns}
\;\;,
\end{displaymath}
so that
\begin{equation}
   \Amat_{n,m}
    =  \weight_{n}^2\delta_{n,m}
      -\weight_{n}\Umeas{\ufunc_{n}\sqrt{\Pdns}\,}
       \weight_{m}\Umeas{\ufunc_{m}\sqrt{\Pdns}\,}
\;\;.
\label{Amat}\end{equation}

\subsection{Gaussian limits}
Without loss of generality, we may assume that the weights $\weight_{n}$ are
ordered from large to small. Then, it is not difficult to see
\cite{vanHameren2001} that the eigenvalues
$\sequence{\lambda_{n}}{n}{\nfunc}$ of the matrix (\ref{Amat}) satisfy
\begin{equation}
   \weight_{1}\geq\lambda_{1}\geq\weight_{2}\geq\lambda_{2}\geq
   \weight_{3}\geq\lambda_{3}\geq\cdots
   \geq\weight_{\nfunc-1}\geq\lambda_{\nfunc-1}
   \geq\weight_{\nfunc}\geq\lambda_{\nfunc}
\;\;.
\label{ordering}\end{equation}
%
%
It is important to realize that (\ref{ordering}) holds whatever $\Pdns$ is.
The influence $\Pdns$ may have on the shape of $\Ptest_{\infty}$ is restricted
to the freedom each of the eigenvalues $\lambda_{n}$ has to change value
between $\weight_{n}$ and $\weight_{n+1}$. The smallest eigenvalue is 
non-negative since the matrix $\Amat$ is positive: for any vector $x$ we
have
\begin{eqnarray}
   \sum_{n,m=1}^{\nfunc}\Amat_{n,m}x_{n}x_{m}
 &=& \sum_{n=1}^{\nfunc}\weight_{n}^2x_{n}^2
    -\bigg(\,\sum_{n=1}^{\nfunc}\weight_{n}
                         \Umeas{\ufunc_{n}\sqrt{\Pdns}\,}x_{n}\,\bigg)^2
\nonumber\\
 &\geq& \sum_{n=1}^{\nfunc}\weight_{n}^2x_{n}^2
    -\bigg(\,\sum_{n=1}^{\nfunc}\weight_{n}^2x_{n}^2\,\bigg)
     \bigg(\,\sum_{n=1}^{\nfunc}\Umeas{\ufunc_{n}\sqrt{\Pdns}\,}^2\,\bigg)
   \geq 0
\;\;,
\nonumber\end{eqnarray}
where the first inequality is by Schwarz, and the second one is based on the
assumption that $\sequence{\ufunc_{n}}{n}{\nfunc}$ is an orthonormal (but not
necessarily complete) set and $\Umeas{\Pdns} = 1$.

For the case that $\Pdns=1$,
it has been shown in \cite{vanHamerenKleissHoogland} that $\Ptest_{\infty}$
becomes Gaussian if and only if there is a limit for the statistic such that
\begin{equation}
   \frac{\lambda_{1}^2}{\sum_{n=1}^{\nfunc}\lambda_{n}^2} \to 0
\;\;.
\label{Gaussianlimit}\end{equation}
Typically, this limit may be $\ndim\to\infty$, as is
shown in various examples. For simplicity, we assume that
$\weight_{1}=\weight_{2}$, which is actually the case in most examples in
\cite{vanHamerenKleissHoogland}. Using this and (\ref{ordering}), it is 
easy to see that, if (\ref{Gaussianlimit}) holds, then also
$\lambda_{1}^2/(\sum_{n=1}^{\nfunc}\weight_{n}^2)\to 0$ and
$\lambda_{1}^2/(\sum_{n=2}^{\nfunc}\weight_{n}^2)\to 0$,
and that the limit holds for any $\Pdns$. So we may conclude that whenever the
spectral decomposition is chosen such that $\weight_{1}=\weight_{2}$ and there
is a limit such that
\begin{displaymath}
  \frac{\weight_{1}^2}{\sum_{n=1}^{\nfunc}\weight_{n}^2} \to 0
\;\;,
\end{displaymath}
then $\Ptest_{\infty}$ becomes Gaussian in this limit.

\section{Example\label{examples}}
The following example of a \gof\ statistic in many dimensions is based on the
diaphony \cite{HellakalekNiederreiter,vanHameren2001}, and has the following
spectral decomposition. The basis is the Fourier basis in $\ndim$ dimensions:
\begin{displaymath}
   \ufunc_{\vec{n}}(\dat) = \prod_{k=1}^{\ndim}\ufunc_{n_k}(\,X_{k}(\dat)\,)
\quad,\quad
   n_{k} = 0,1,2,\ldots
\quad,\quad
   k=1,2,\ldots,\ndim
\;\;,
\end{displaymath}
with
\begin{displaymath}
   \ufunc_{0}(x) = 1
   \quad,\quad
   \ufunc_{2n-1}(x) = \sqrt{2}\sin(2\pi nx)
   \quad,\quad
   \ufunc_{2n}(x) = \sqrt{2}\cos(2\pi nx)
\;\;,
\end{displaymath}
for $n$ from $1$ to $\infty$. The corresponding weights are given by
\begin{displaymath}
   \weight_{\vec{n}} = \prod_{k=1}^{\ndim}\weight_{n_k}
   \qquad\textrm{with}\qquad
   \weight_{0} = 1
   \quad,\quad
   \weight_{2n-1} = \weight_{2n} = \frac{1}{n}
\;\;.
\end{displaymath}
The two-point function is equal to
\begin{displaymath}
   \cmu(\dat_{1},\dat_{2})
   = \sum_{\vec{n}}
     \weight_{\vec{n}}^2\ufunc_{\vec{n}}(\dat_{1})\ufunc_{\vec{n}}(\dat_{2})
   = \prod_{k=1}^{\ndim}\cmu_{1}(\,X_k(\dat_{1})-X_k(\dat_{2})\,)
\;\;,
\end{displaymath}
where 
$\sum_{\vec{n}}=\sum_{n_{1}=0}^{\infty}\sum_{n_{2}=0}^{\infty}
                \cdots\sum_{n_{\ndim}=0}^{\infty}$
and
\begin{displaymath}
   \cmu_{1}(x)
   = 1+\frac{\pi^2}{3}-2\pi^2(x\,\mod\,1)(1-x\,\mod\,1)
\;\;.
\end{displaymath}
The only important difference with the two-point function of the diaphony is
that there the constant mode, the $\ndim$-dimensional basis function which is
equal to $1$, is missing. This makes sense since the diaphony is constructed in
order to test the uniform distribution and the contribution of the constant
mode cancels in (\ref{defDisc}). The advantage is that the diaphony is directly
given by the sum of all two-point correlations between the data-points and no
integrals of two-point functions have to be calculated. Notice that this
cancellation also appears in (\ref{Amat}): the first row and column of the
matrix $\Amat$ consist of only zeros if $\Pdns=1$, since all modes except the
constant mode have zero integral. For a general \pdf\, these cancellations also
exist, but not for a single mode, and hence are not of practical use. For
example, $\ffunc=1/\Qdns$ cancels in (\ref{defgof}).

\begin{figure}
\begin{center}
\epsfig{figure=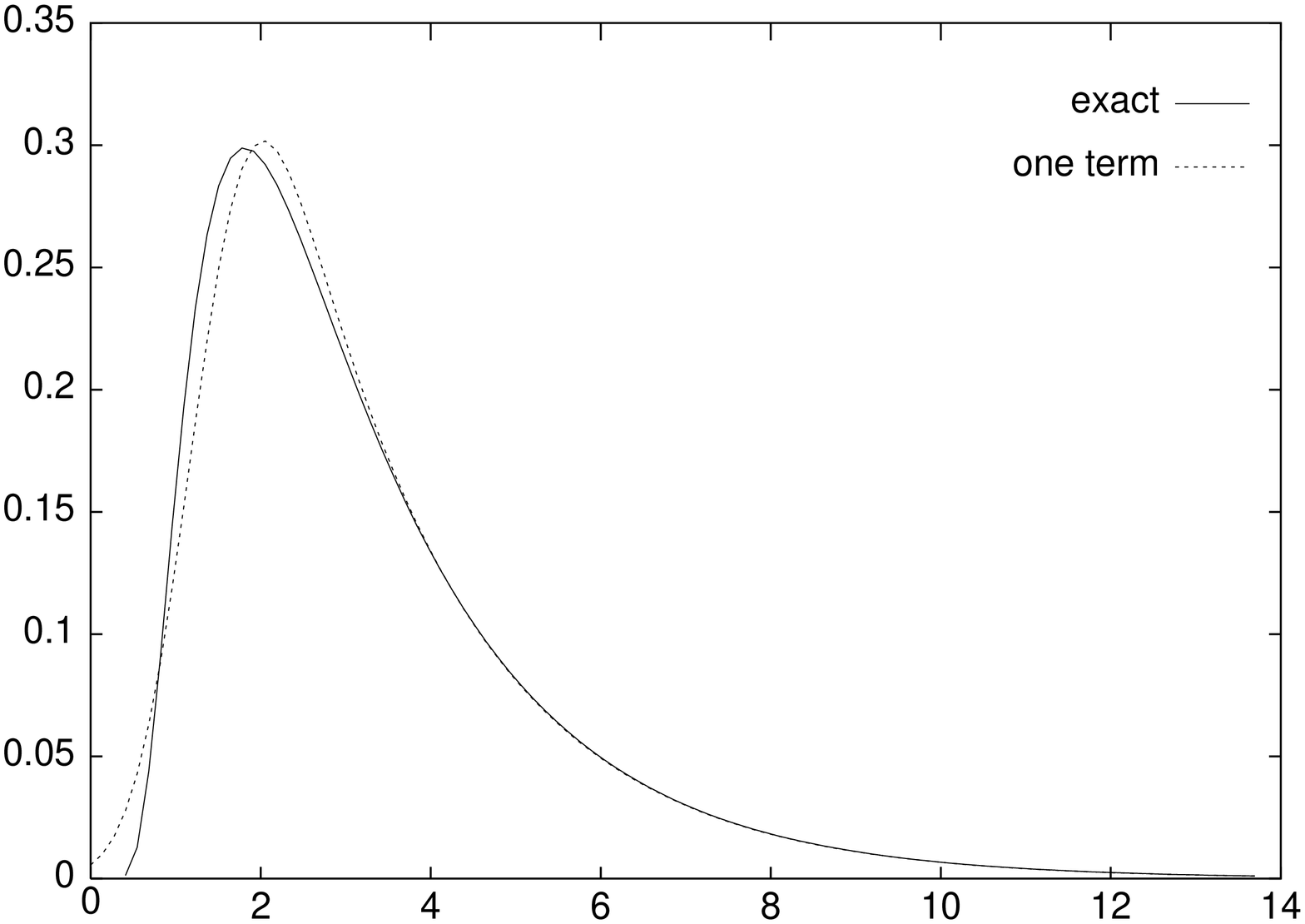,width=0.5\linewidth,angle=270}
\caption{$\Ptest_{\infty}(t)$ for $\ndim=1$ and $\Pdns=1$ using (\ref{Gen1dim})
and (\ref{logGen}) with one term.}
\label{fig01}
\end{center}
\end{figure}

\begin{figure}
\begin{center}
\epsfig{figure=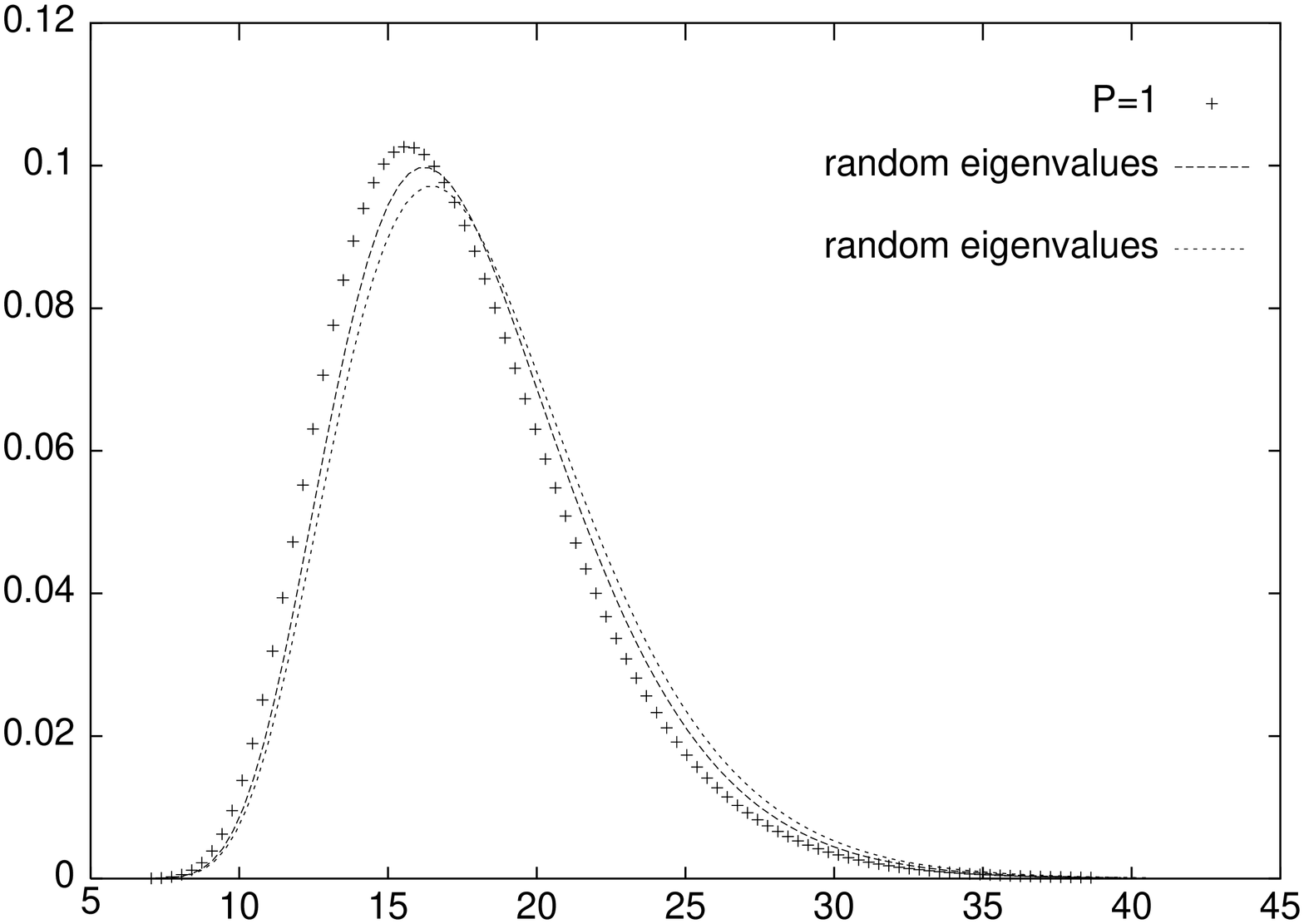,width=0.5\linewidth,angle=270}
\caption{$\Ptest_{\infty}(t)$ for $\ndim=2$, for $\Pdns=1$ and two sets of
random eigenvalues $\lambda_{s}$.}
\label{fig02}
\end{center}
\end{figure}

\begin{figure}
\begin{center}
\epsfig{figure=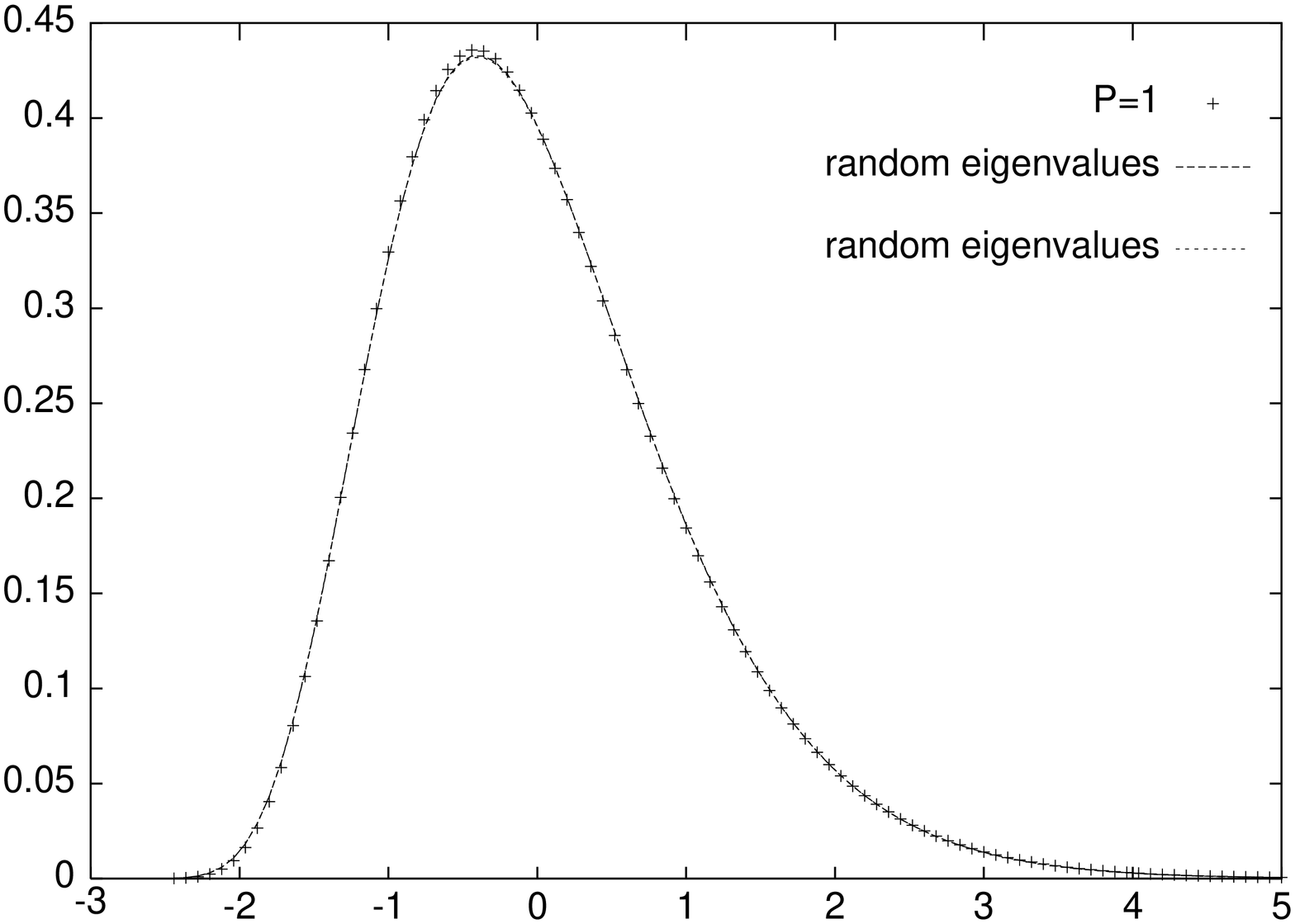,width=0.5\linewidth,angle=270}
\caption{$\sqrt{\VV}\,\Ptest_{\infty}(\sqrt{\VV}\,t+\EE)$ for the same
situations as in \Figure{fig02}.}
\label{fig03}
\end{center}
\end{figure}

It is useful to introduce the
function $\nequal$ which counts the number of weights with the same value:
\begin{displaymath}
   \nequal(s)
   = \sum_{\vec{n}}\delta_{s,1/\weight_{\vec{n}}}
\;\;.
\end{displaymath}
The numbers $\nequal(s)$ increase as function of $\ndim$.
Using $\nequal$, (\ref{Geninfty1}) and (\ref{ordering}), the generating
function can be written as
\begin{displaymath}
   \Gen_{\infty}(z)
   = \bigg(\;\prod_{s=1}^{\infty}(1-2z/s^2)^{\nequal(s)-1}
                                 (1-2z\lambda_{s})\;\bigg)^{-1/2}
\;\;,
\end{displaymath}
where the numbers $\lambda_{s}$ depend on the \pdf\ under consideration,
but are, following (\ref{ordering}), restricted by the relation
\begin{displaymath}
   1/s^2>\lambda_{s}\geq 1/(s+1)^2
\;\;.
\end{displaymath}
In order to find the probability density $\Ptest_{\infty}$, the inverse Laplace
transformation (\ref{Laplace}) has to be performed on $\Gen_{\infty}$. The
logarithm of the product can best be evaluated as described in
\cite{JamesHooglandKleiss}, by extracting the first and the second order terms
in $z$:
\begin{equation}
   \log\Gen_{\infty}(z)
   = \EE z + \srac{1}{2}\VV z^2
   + \sum_{s=1}^{\infty}(\,g_{s}(z)-g_{s}'(0)z-\srac{1}{2}g_{s}''(0)z^2\,)
\;\;,
\label{logGen}\end{equation}
where
$g_{s}(z)=-\srac{\nequal(s)-1}{2}\log(1-2z/s^2)
          -\srac{1}{2}\log(1-2z\lambda_{s})$,
and $\EE$ and $\VV$ are the expectation value and the variance of the
statistic. For the case that $\Pdns=1$, so that $\lambda_{s}=1/(s+1)^2$, they
can be calculated directly and are given by
\begin{displaymath}
   \EE_{\mathrm{uniform}} = \bigg(1+\frac{\pi^2}{3}\,\bigg)^{\ndim}-1
\quad,\quad
   \VV_{\mathrm{uniform}} = 2\bigg(1 + \frac{\pi^4}{45}\,\bigg)^{\ndim}-2
\;\;.
\end{displaymath}
We want to study the influence of $\Pdns$ on $\Ptest_{\infty}$ by generating
the eigenvalues $\lambda_{s}$ at random, uniformly distributed within their
borders, and plotting the result. First, however, we need to find out how many
terms in the infinite sum of (\ref{logGen}) have to be taken into account in
order to obtain a trustworthy result. This can be done at $\ndim=1$, since we
know already that $\Ptest_{\infty}$ will tend to look like a Gaussian for
larger values of $\ndim$ so that the sum must become less important.
Furthermore, there is the advantage that at $\ndim=1$ and $\Pdns=1$ there
exists a simple formula for the generating function:
\begin{equation}
   \Gen_{\infty}^{\mathrm{uniform}}(z) 
   = \frac{\sqrt{2\pi^2 z}}{\sin\sqrt{2\pi^2 z}}
\;\;.
\label{Gen1dim}\end{equation}
In \Figure{fig01}, we present the result with this formula and with
(\ref{logGen}) using only one term. With $10$ terms, the difference between the
curves is invisible and this is the number we further use.
Results for $\ndim=2$ are depicted in \Figure{fig02}. As expected, the curves
look more `Gaussian' than the $1$-dimensional curve. The crosses represent the
case $\Pdns=1$, and the two continuous curves represent two cases with 
`typpical' sets of random eigenvalues. The curves are clearly different, but if
we go over to standardized variables, that is, if we plot
\begin{displaymath}
   \sqrt{\VV}\,\Ptest_{\infty}(\sqrt{\VV}\,t+\EE)
\;\;,
\end{displaymath}
so that the expectation value is equal to $0$ and the variance is equal to $1$,
we find \Figure{fig03}, and we may conclude that the curves almost only depend
on the expectation value and the variance.  Again, we know that this behavior
becomes only stronger for higher values of
$\ndim$ because of the Gaussian limit.
We conclude that $\Pdns_{\infty}$ for general $\Pdns$ can, to satisfying
accuracy, be approximated by
\begin{displaymath}
   \sqrt{\VV_{\mathrm{uniform}}/\VV}
 \;\Ptest_{\infty}^{\mathrm{uniform}}\Big(
    \sqrt{\VV_{\mathrm{uniform}}/\VV}\,(t-\EE)+\EE_{\mathrm{uniform}}\Big)
\;\;,
\end{displaymath}
where $\Ptest_{\infty}^{\mathrm{uniform}}$, $\EE_{\mathrm{uniform}}$ and
$\VV_{\mathrm{uniform}}$ are the asymptotic test-distribution, the expectation
value and the variance for the case that $\Pdns=1$.

\section{Conclusion}
We have seen how to construct practical \gof\ statistics to test the hypothesis
that a sample of data is distributed following a given \pdf\ in many
dimensions, for which the asymptotic test-distribution in the limit of an
infinite sample size becomes Gaussian in the limit of an infinite number of
dimensions.
Furthermore, we have seen an explicit example of such a statistic, for which
the asymptotic test-distribution depends on the \pdf\ as good as only through
the expectation value and the variance of the statistic for any number of
dimensions larger than one.

\subsubsection*{Acknowledgment}
This research has been supported by a Marie Curie Fellowship of the European
Community program ``Improving Human Research Potential and the Socio-economic
Knowledge base'' under contract number HPMD-CT-2001-00105,
and Deutche Forschungsgemeinschaft through the Graduiertenkolleg
`Eichtheorien' at the Johannes-Gutenberg-Universit\"at, Mainz.

\newcommand{\EPJ}[3]{Eur.\ Phys.\ J.\ {\bf #1} (#2) #3}
\newcommand{\CPC}[3]{Comp.\ Phys.\ Comm.\ {\bf #1} (#2) #3}
\newcommand{\JCP}[3]{J.\ Comp.\ Phys.\ {\bf #1} (#2) #3}
\newcommand{\PL}[3]{Phys.\ Lett.\ {\bf #1} (#2) #3}
\newcommand{\NP}[4]{Nucl.\ Phys.\ {\bf #1} #4 (#2) #3}
\newcommand{\lb}{\linebreak[1]}


\begin{thebibliography}{}
\bibitem{AslanZech2002}
B.~Aslan and G.~Zech,
{\it Comparison of Different Goodness-of-Fit Tests\/},
Durham 2002, Advanced statistical techniques in particle physics 166-175
({\tt http://\lb www.\lb ippp.\lb dur.\lb ac.\lb uk/\lb Workshops/\lb 02/\lb statistics/\lb proceedings.\lb shtml}).
\bibitem{AslanZech2002-2}
B.~Aslan and G.~Zech,
{\it A new class of binning free, multivariate goodness of fit tests: the
energy tests\/}
({\tt http://arxiv.org/abs/hep-ex/0203010}).
\bibitem{Raja2003}
R.~Raja,
{\it A Measure of the Goodness of Fit in Unbinned Likelihood Fits; 
     End of Bayesianism},
eConf C030908:MOcT003, 2003
({\tt http://\lb arxiv.org/\lb abs/\lb physics/\lb 0401133}).
\bibitem{Abraham1999}%
K.J.~Abraham,
{\it A New Technique for Sampling Multi-Modal Distributions\/}
({\tt http://\lb arxiv.\lb org/\lb abs/\lb physics/\lb 9903044}).
\bibitem{Tichy1997}
R.F.~Tichy and M.~Drmota,
{\it Sequences, Discrepancies and Applications\/}
(Springer, 1997).
\bibitem{Niederreiter1992}
H.~Niederreiter,
{\it Random number generations and Quasi-Monte Carlo methods\/}
(SIAM 1992).
\bibitem{QuasiMonteCarlo}
{\it Monte Carlo \& Quasi-Monte Carlo Methods\/}
({\tt http://www.mcqmc.org}).
\bibitem{vanHamerenKleissHoogland}
A.\ van Hameren, R.\ Kleiss and J.\ Hoogland,
{\it Gaussian limits for discrepancies: I.\ Asymptotic results\/},
\CPC{107}{1997}{1-20} 
({\tt http://\lb arxiv.\lb org/\lb abs/\lb physics/\lb 9708014}).
\bibitem{vanHamerenKleiss}
A.\ van Hameren and R.\ Kleiss,
{\it Quantum field theory for discrepancies\/},
\NP{B529}{1998}{737-762}{[PM]} 
({\tt http://\lb arxiv.\lb org/\lb abs/\lb math-ph/\lb 9805008}).
\bibitem{HellakalekNiederreiter}
P.~Hellakalek and H.~Niederreiter,
{\it The weighted spectral test: Diaphony\/},
ACM Trans.\ Model.\ Comput.\ Simul.\ 8, No.\ 1 (1998), 43-60.
\bibitem{vanHameren2001}
A.~van~Hameren,
{\it Loaded Dice in Monte Carlo: Importance sampling in phase space integration 
     and probability distributions for discrepancies\/},
PhD-thesis (Nijmegen, 2001)
({\tt http://\lb arxiv.\lb org/\lb abs/\lb hep-ph/\lb 0101094}).
\bibitem{JamesHooglandKleiss}
F.~James, J.~Hoogland and R.~Kleiss,
{\it Multidimensional sampling for integration and simulation:
     measures, discrepancies and quasi-random numbers\/},
\CPC{99}{1997}{180-220}
({\tt http://\lb arxiv.org/\lb abs/\lb physics/\lb 9606309}).
\end{thebibliography}
\end{document}